\documentclass[twocolumn,showpacs,preprintnumbers,amsmath,amssymb]{revtex4}
\usepackage{graphicx}

\begin{document}
\title{Pattern formation and  nonlocal logistic growth}
\author{Nadav M. Shnerb}
\affiliation{Department of Physics, Bar-Ilan University, Ramat-Gan
52900 Israel}
\begin{abstract}
Logistic growth process with nonlocal interactions is considered
in one dimension. Spontaneous breakdown of translational
invariance is shown to take place at some   parameter region, and
the bifurcation regime is identified for short and long range
interactions. Domain walls between regions of different order
parameter are expressed as soliton solutions of the reduced
dynamics for nearest neighbor interactions. The analytic results
are confirmed by numerical simulations.
\end{abstract}


 \maketitle

\section{Introduction}
\noindent Logistic growth is one of the basic models in population
dynamics. First introduced by Verhulst for saturated proliferation
at a single site, it has been extended to include spatial dynamics
by Fisher \cite{fisher} and by Kolmogoroff et. al. \cite{kol}. In
its one-dimensional  continuum version, one consider  the
concentration of a reactant, $c(x,t)$, with time evolution  that
is given by the rate equation:
\begin{equation}
\frac{\partial c(x,t)}{\partial t}= D \nabla^2
c(x,t)+ac(x,t)-bc^2(x,t),
\end{equation}
where $D$ is the diffusion constant, $a$ is the growth rate and
$b$ is the  saturation coefficient set by the carrying capacity of
the medium.

The Fisher process  is a generic description of the invasion of a
stable phase into an unstable region. It is applicable to  wide
range of phenomena, ranging from genetics (the original context of
Fisher work, proliferation of a favored mutation or gene) to
population dynamics, chemical reactions in unstirred reactors,
hydrodynamic instabilities, invasion of normal states by
superconducting front, spinodal decomposition and many other
branches of natural sciences. A comprehensive survey may be found
in recent review article by van-Saarloos \cite{saarloos}.

The  Fisher process ends up with a uniform saturated phase, in
contrast with  other nonlinear and reactive systems that yields
spatial structures with no underlying inhomogeneity. These
patterns are usually related  to an instability of the homogenous
solution, most commonly of Turing or Hopf types \cite{book}.
Spontaneous symmetry breaking of that type manifests itself in
vegetation patterns, where competition of flora for common
resource (water) induces an indirect interaction  and may lead to
a (Turing like)  spatial segregation \cite{merron}.

The basic motivation of this work comes from recent study of
non-Turing mechanism for pattern formation in the vegetation-water
system, that yields ordered or  glassy structures \cite{shnerb}.
Basically, it is easy to realize that \emph{competition for common
resource induced some indirect "repulsion" among agents, that may
lead to spatial segregation}. As an example, consider the
vegetation case: there is a constant flow of water into the system
(rain), and the water dynamics (evaporation, percolation,
diffusion) is much faster than the dynamics of the flora. Now let
us assume the existence of a large amount of flora (say, a tree)
at certain spatial point. One may expect the water density to
adjust (almost instantly) to the tree and to equilibrate in some
water profile that is lower around its location. The immediate
neighborhood of the tree, though, is less favorable for a second
tree to flourish; instead one may expect the next to grow up some
typical distance away, reducing the water level between them even
more. This seems to be a plausible and generic  mechanism for
segregation induced by resource competition. These arguments may
be relevant to the dynamics of almost any unstirred reactive
system; interesting example is the process of evolutionary
\emph{speciation}, where new species may survive only "far enough"
(in the genome space, where the spatial structure is given, say,
by Hamming distance) from its ancestor, in order to find a
non-overlaping biological niche.

Surprisingly it turns out that the partial differential equations
that describe this process (here presented in a nondimensionalized
form, where  $w$ stands for water density, $b$ for flora and $R$
is the "rain"):
\begin{eqnarray}
{\dot b(x,t)}&=& \nabla^2 b- \mu b-  w b \nonumber
\\{\dot w(x,t)}&=& D\nabla^2 w +R -w- w b,
\end{eqnarray}
yield only a linearly \emph{stable} homogenous solution. In order
to get patterns one should add a cross-diffusion effect (slowing
down of the water diffusion in the presence of flora) that leads
to Turing-like instability as in \cite{merron}, but this is a
different mechanism, and one may wander about the validity of the
basic intuitive argument presented above.

Recent work \cite{shnerb} suggests a hint for the answer. It seems
that a continuum and local description of a reactive system fails
to capture the competition induced segregation discussed above.
The continuum process is trying to "smear" the reactant profile,
and instead of getting spatially segregated structure of large
biomass units ("trees") it favors homogenous profile of "grass"
covering all the area. In \cite{shnerb} a biomass unit was allowed
for long time survival  only if it exceeds some pre-determined
threshold, and simulation of the system reveal an immediate
appearance of spontaneous segregation and stable patterns.

Similar situation appears, presumably, in the process of bacterial
colony growth where the food supply is limited. As noted by
Ben-Jacob et. al. \cite{eshel}, spatial segregation and branching
are induced by the competition of bacteria for diffusive food. A
communicating walkers model is used by these authors to simulate
the branching on a petri-dish, where  continuum equation dictates
the food dynamics while the individual bacteria are discrete
objects. The discreteness of bacteria adds some weak threshold to
the system and induces segregation. Note, however, that the model
admits weak dependence of the diffusion on the local bacterial
density at the boundaries of the colony.

In this work I consider the one-species analogy of the competition
problem, namely, a logistic growth with \emph{nonlocal
interactions}, where the carrying capacity at a site is reduced
due to the presence of "life" in another site. Nonlocal
competition has been recently considered by Sayama et al.
\cite{Sayama} and by Fuentes et al. \cite{Fuentes}. Both groups
uncover the possibility of spontaneous symmetry breaking and
patterns, depending on the the strength and the smoothness of the
"weight function" that controls the nonlocality. It seems that
nonlocal interactions are not simply an effective model obtained
by integrating out the fast degrees of freedom; rather, it
incorporates some nonlinear effects (like the threshold) and allow
for linear instability that manifests the intuitive "competition
induced segregation" argument.

Sayama et al. \cite{Sayama} deal with a two dimensional model of
population dynamics, with no diffusion term. Both the local growth
term  and  the carrying  capacity at a site depend (not in the
same way) on the population of  neighboring sites; in a crowded
neighborhood the growth term becomes larger (due to offspring
migration) while the carrying capacity decreases as a result of
long range competition. The conditions for an instability of the
homogenous solution have been found analytically and demonstrated
numerically for a "stepwise" weight function (taking as the
effective neighborhood the average density inside a prescribed
radius around the site). It was also pointed out that a gaussian
weight function yields no instability.

In the numerical work of \cite{Fuentes}, a one dimensional
realization of diffusing reactants has been considered, equivalent
to Fisher equation with non-local interactions. Again it was shown
that a stepwise weight function may lead to instability while a
gaussian weights lead to stable homogenous solution; the authors
proceed to consider intermediate weight functions that interpolate
between gaussian and a step function.

As in any case of spontaneous symmetry breaking, the system falls
locally into one of the "minima" of the order parameter, and
typically domains are formed. These domain walls determine the low
lying excitation spectrum of the system, as their movement is
"smooth": if the broken symmetry is continuous the resulting
Goldstone modes may destroy the long range order at finite
temperature, and the same is true for the domain walls if discrete
symmetry is broken. Although we are dealing with an out of
equilibrium system, one may guess that the response to small noise
is determined by these domain walls.

The goals of this work are twofold: in the next section I will try
to give more comprehensive discussion of the instability
condition, with and without diffusion, and its dependence on the
weight functions: it turns out that it depends on the minimal
value of its Fourier transform. The third section  is devoted to
the appearance of topological defects in the segregated phase.
Finally in section IV some discussion and possible implications
are presented.

\section{Instability conditions}
\noindent The model is a  one dimensional realization of
long-range competition system  on a lattice (with lattice spacing
$l_0$) and the continuum limit is trivailly attained at $l_0 \to
0$.

In the generic case of diffusion and non-locality  the time
evolution of the reactant density at the n-th site,
$\widetilde{c}_n$, is given by:

\begin{eqnarray}
\label{eq:1} \frac{\partial \widetilde{c}_n(t)}{\partial
t}&=&\frac{\widetilde{D}}{l_0^2}[-2\widetilde{c}_n(t)+\widetilde{c}_{n+1}(t)+\widetilde{c}_{n-1}(t)]+
a \widetilde{c}_n(t)  \nonumber \\  &-&b \widetilde{c}_n^{2}(t) -
\widetilde{c}_n(t) \sum_{r=1}^\infty \widetilde{\gamma}_r
[\widetilde{ c}_{n+r}(t)+\widetilde{c}_{n-r}(t)].
\end{eqnarray}
where $\widetilde{D}$ is the diffusion constant and
$a,b,\widetilde{\gamma}$ are the corresponding reaction rates. The
definition of dimensionless quantities,
\begin{equation} \tau=at, \qquad
 c=b\widetilde{c}/a, \qquad  \gamma_r=\widetilde{\gamma}_r/b,
 \qquad
  D=\frac{\widetilde{D}}{a l_0^2}.
  \end{equation}
(the new   "diffusion constant" is $D = W^2/l_0^2$, where  $W
\equiv \sqrt{D/a}$ is the width   of the Fisher front) provides
the dimensionless equation,
\begin{eqnarray}
\label{eq:2} \frac{\partial c_n}{\partial \tau}&=& D
[-2c_n+c_{n+1}+c_{n-1}] \nonumber \\ &+& c_n \left( 1-c_n-
\sum_{r=1}^\infty \gamma_r [c_{n+r}+ c_{n-r}] \right),
\end{eqnarray}
that may be expressed in Fourier space  [with $A_k \equiv \sum_n
c_n e^{iknl_0}$] as,
\begin{equation} \label{eq:3}
\dot{A}_k = \alpha_k A_k - \sum_q \beta_{k-q} A_q A_{k-q},
\end{equation}
where
\begin{eqnarray}
\alpha_k \equiv 1-2D[1-cos(kl_0)] \\
\beta_k \equiv 1+2\sum_{r=1}^\infty \gamma_r cos(rkl_0).
\end{eqnarray}
As $c_n$ is positive semi-definite, $A_0$ is always "macroscopic".
Any mode is  suppressed by $A_0$, and for small $\gamma_r$ one
expects only the zero mode to survives \cite{ns}. If, on the other
hand, $\gamma_r$ increased above some threshold, bifurcation may
occur with  the activization of some other $k$ mode(s), and the
homogenous solution becomes unstable. This is the situation where
patterns appear and translational symmetry breaks.

To get a basic insight into the problem, let us consider the case
with \textbf{no diffusion} ($D=0$, $\alpha_k=1$). Eq.(\ref{eq:2})
becomes,
\begin{equation} \label{nodiff}
\dot{c}_n = c_n \left[ 1-c_n -\sum_r \gamma_r (c_{n+r} - c_{n-r})
\right]
\end{equation}
and division by $c_n$ yields, for the steady state, the linear
equation $\mathcal{Q}.\underline{c}=\underline{y}$, where
$\mathcal{Q}$ is a circular matrix, $\underline{c}$ is the vector
of $c_n$-s  and $\underline{y}=(...1,1,1,1,...)^{\dag}$.  The sum
of the elements of any row of $\mathcal{Q}$ is the same, so the
homogenous state (scalar multiplication of $\underline{y}$) should
be an eigenvector. On the other hand,  if $\mathcal{Q}$ is
non-singular it must admit a full set of mutually orthogonal
eigenvectors. Only the constant eigenvector of $Q$ has
nonvanishing projection on $y$, so the only \emph{positive
definite}, non diverging steady state ($\dot{c_n}=0 \quad c_n
>0 \quad \forall{n}$) solution is the \emph{homogenous} one,
$c_n = 1/\beta_0$.

\begin{figure}
  \includegraphics[width=7.7cm]{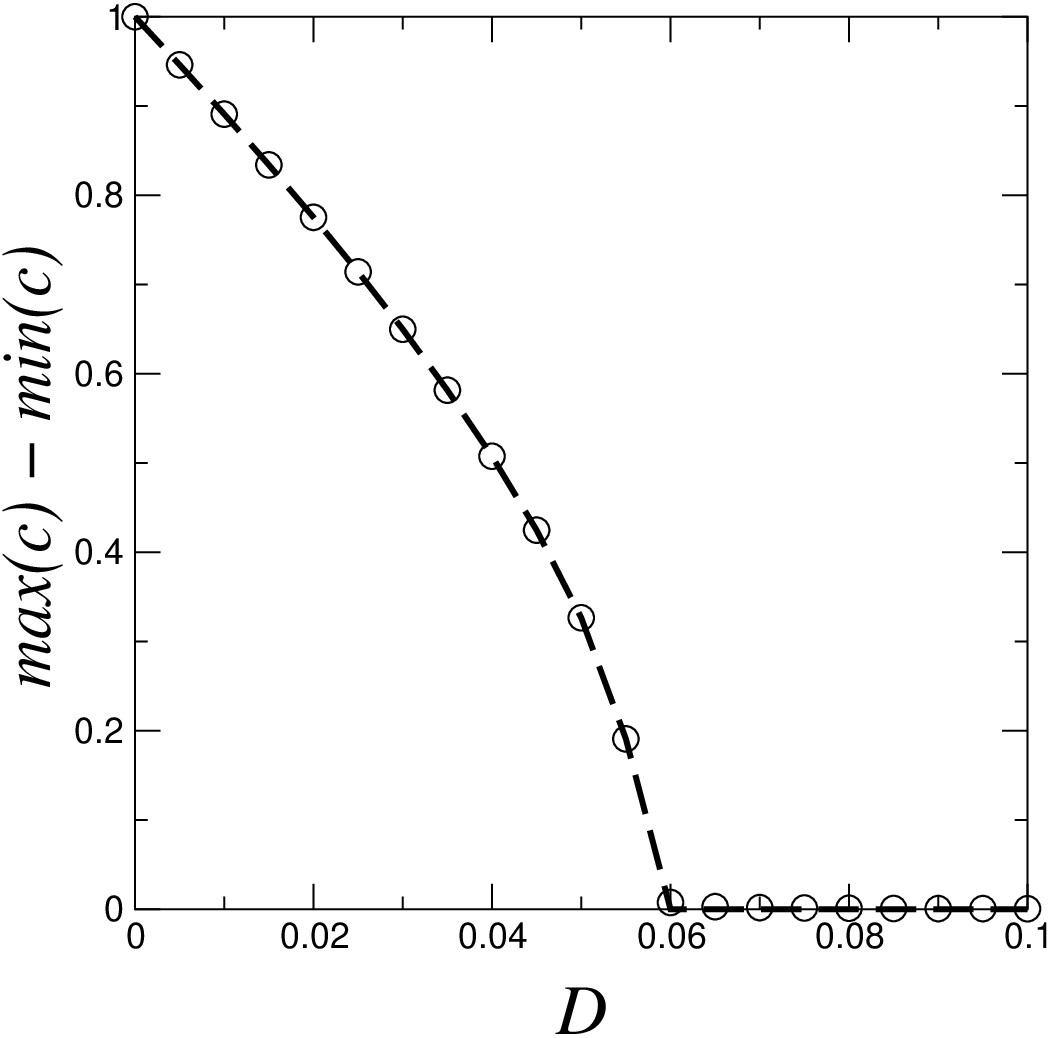}\\
  \caption{Maximal amplitude of  $c_n$ differences [$max(c_n)-min(c_n)$](circles) for a sample of 1024 sites
  (periodic boundary conditions)  and the predicted difference according to
  Eq. (\ref{diff}) for nearest neighbor interaction with $\gamma = 0.8$. (dashed
  line).
  The agreement is up to the numerical error.}\label{fig1}
\end{figure}

As implied by Eq. (\ref{eq:3}), the homogenous steady state is
unstable iff, for some $k$, $\beta_k < 0$. In that case
bifurcation occurs, and the new steady state is a combination of
the zero mode and the $k$ mode with equal weights $A_0 = A_k =
1/(\beta_0 + \beta_k)$.

\begin{figure}
  \includegraphics[width=7.7cm]{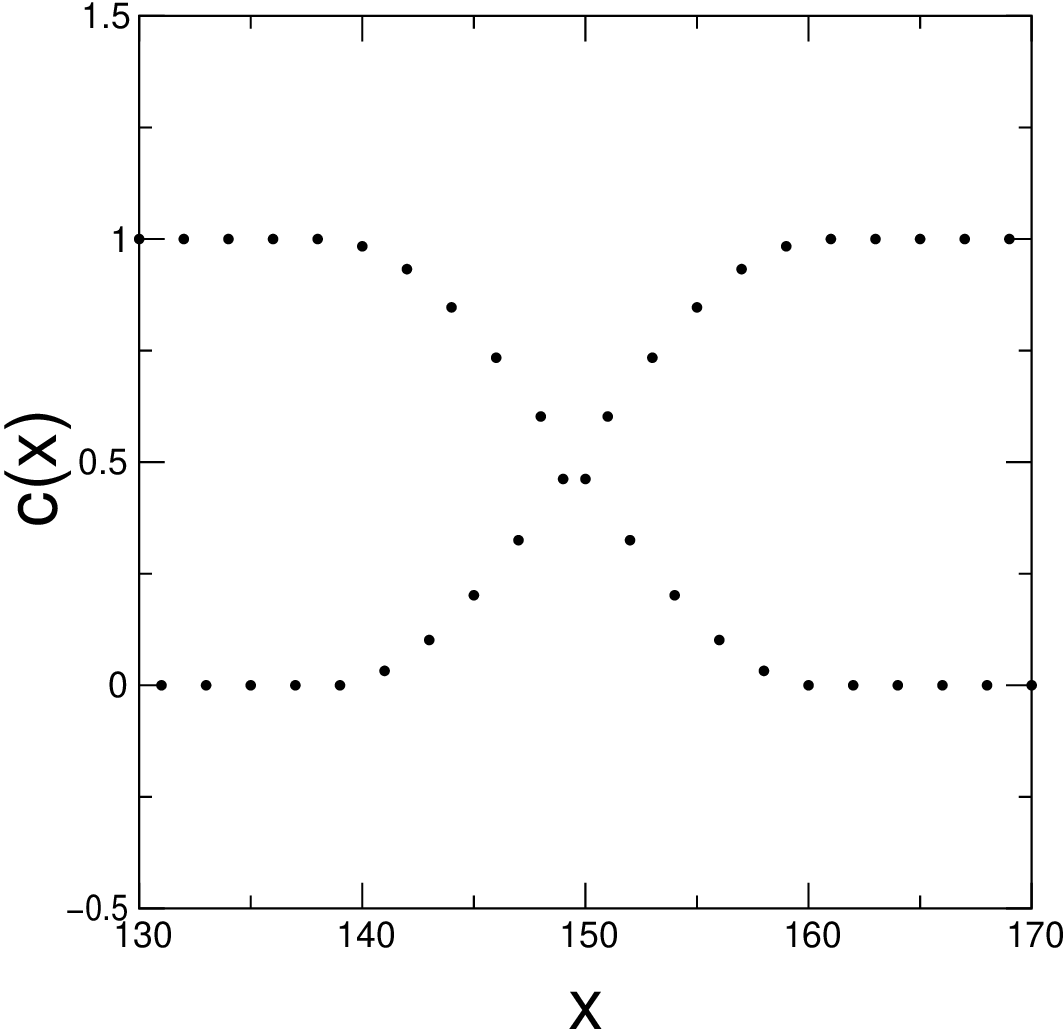}\\
  \caption{A typical "soliton" of length $L=20$, an outcome of forward Euler integration of Eq. (\ref{eq:6})
  on 1024 lattice points with periodic boundary conditions and random initial conditions at
   $\gamma = 0.505$ (just above the bifurcation). There is a perfect agreement
   with the theoretical prediction,
   Eq. (\ref{soliton}),  up to the accuracy of the numerics (here, 4-5 significant digits)} \label{fig2}
\end{figure}

The function $\beta_k, \  k \in [0,\pi/l_0]$, is discrete for
finite systems and becomes continuous at the thermodynamic limit.
If $\beta_k$ never crosses zero there is no bifurcation and the
homogenous solution $c_n = 1/\beta_0$ is stable. The results for
few types of interaction ranges, with the critical value
$\gamma_1^c$  (where the instability occurs), and $k_c$ (the first
excited mode), are summarized in Table I.

\begin{table*}\label{table}
\begin{large}
  \centering

\begin{tabular}{|c|c|c|c|}
  \hline
 type & $\gamma_r$ & $\beta_k $ & $\rm{Instability \  condition}$ \\
  \hline
exponential & $ (\gamma_1)^{r/\xi} $ &  $
\frac{sinh(|ln(\gamma_1)|/\xi)}{cosh(|ln(\gamma_1)|/\xi)-\cos(kl_0)}$
 & $\rm{no  \ instability}$ \\
quadratic &  $\frac{\gamma_1}{r^2}$  & $1 + 2 \gamma_1
[\frac{\pi^2}{6}-\frac{\pi k l_0}{2} + \frac{(kl_0)^2}{4}]$ &
 $\gamma_1^c = \frac{6}{\pi^2}$,  $k_c = \pi/l_0$ \\

step & $\gamma_r = \{\begin{array}{cc}
  1 & r \leq p \\
  0 & r > p  \\
\end{array} $ & $ \frac{sin(\frac{ka}{2} (2p+1))}{sin(\frac{ka}{2})}$ & at large p $k_c = \frac{4 p a}{3}$ \\

gaussian & $ (\gamma_1)^{(r^2/\sigma^2)} $ &  if $\sqrt{\sigma} >>
l_0 $ , $\sim \frac{\sigma}{2} \sqrt{\frac{\pi}{|ln(\gamma_1)|}}
exp \left( \frac{(\sigma k l_0)^2}
{4  \sqrt{|ln(\gamma_1)|}} \right)$ & $\rm{no  \ instability}$ \\

  \hline
\end{tabular}
\end{large}
\caption{The function $\beta_k$ and the instability conditions for
various types of  nonlocal interactions. The results for the
Gaussian case are in the continuum approximation.}
\end{table*}


It is interesting to note that these expressions may be
generalized to yield a full,  period doubling type, instability
cascade. The m-th instability involves $2^m$ modes, and the steady
state is $1/\sum_k \beta_k$, where the sum runs over all the
"active" modes. The condition for the $m+1$ bifurcation
[activation of another $2^{m+1}$ modes] is the existence of a
wavenumber $q$ such that $\sum_k \beta_{q-k} <0$, with the sum
runs, again, over all $2^m$ active k-s. There are, however, some
obstacles for the implication of these solutions above the first
bifurcation. Degeneracies in $\beta_k$ (e.g., for $\gamma_r =
\delta_{r,4}$, both $kl_0=\pi/4$ and $kl_0=3\pi/4$ are minima),
and solitons between different stable phases (described below) may
blur the native state. In this letter, though, $\beta_k$ is used
only for the first, pattern-forming, instability criteria, and the
details of the emerged structure are presented just for
nearest-neighbor interaction.

Once diffusion is added to the system, its features changes, but
not so much. The homogenous state is still characterized by $c_n =
1/\beta_0$ and the first pattern formation instability  appears
when some $k$ mode satisfy:
\begin{equation}
\beta_k < - 2 \beta_0 D [1-cos(kl_0)].
\end{equation}
Above this instability, the amplitudes of the modes are not equal,
\begin{equation} \label{diff}
A_0 = \frac{\alpha_k}{\beta_0 + \beta_k} \qquad  A_k =
\sqrt{\frac{\alpha_k(\beta_0 + \beta_k) - \beta_0}{\beta_k
(\beta_0 +\beta_k)^2}}
\end{equation}
and there are no zeroes of $c_n$. This result fits perfectly with
the numerical data presented in  figure (\ref{fig1}). Again the
m-th instability involves the activation of $2^m$ modes, Although
the stability analysis is more complicated.

The question of pattern instability is thus translated to the
determination of the minimal value of the Fourier coefficient of
the weight function (or the "weight series" $\gamma_r$). If the
minimal value is smaller than some prescribed number (zero if
there is no diffusion) instability takes place and patterns
emerge. Unfortunately I am not familiar with a general theorem
that sets bounds on the minimal value of the Fourier coefficient
of a function based on its "smoothness", or other analytic
properties, so any case should be considered separately, with the
generic examples given in Table I.

\section{Domain wall structure}
\noindent Above the pattern formation threshold generic initial
conditions fail to yield perfect "lattice", as different domains
reach saturation with different "phases". These domains are
connected by soliton-like solutions of the time independent
equation in the following sense: any stable solution (${\dot
c}(x,t)=0$) should satisfy (in the  continuum limit ),
\begin{equation}
D \frac{d^2 c(x)}{dx^2} = c(x)  - c(x) \int f(x-y) c(y) dy,
\end{equation}
thus it looks like a trajectory of mechanical particle (with mass
D) in a nonlocal potential, with $x$ as the "time". A domain wall
is a finite size structure, so it must connect  fixed points of
this fictitious dynamics, i.e., a domain wall corresponds to
\emph{heteroclinic orbit}. In this section we consider these
"solitons" and look for their shape and size at different
conditions. In order to simplify the discussion, only the nearest
neighbor case is considered, both with and without diffusion.

With no diffusion and n.n. competition,   (\ref{nodiff}) takes the
form:
\begin{equation}
\label{eq:6} \frac{\partial c_n}{\partial \tau}= c_n\left[
1-c_n-\gamma (c_{n+1}+ c_{n-1}) \right].
\end{equation}
The uniform solution, in this case, is $c=\frac{1}{1+2 \gamma}$,
and the nonuniform solution is, either $c_n = 1$ for odd n and
$c_{n}=0$ for even, or vice versa. Stability analysis shows that
the uniform solution becomes unstable at $\gamma_c=1/2$, and the
zero-one phase is stable above this value. One may expect, though,
to see a jump from homogenous to  patterned  (zero-one) phase at
$\gamma_c$. However,  if the initial conditions are taken from
random distribution, there is a chance for a domain wall between
two regions, as indicated by the numerical results presented in
Figure (\ref{fig2}).

Clearly, such a soliton should be  a solution of the "map"
\begin{equation}
\label{eq:7} c_{n+1}= \frac{1-c_n}{\gamma} - c_{n-1}.
\end{equation}
of course, 01010101... (odd 0-s) and 101010101... (even 0-s) are
already solutions of this equation. We are looking for the
solution that connect these two fixed points. Such a trajectory
begins in, say, 010101010 state, but then after the zero it gives
not 1 but $x_1$. The dynamics now continue in a different
trajectory, but $x_1$ should be selected such that after L steps
of the map (for domain wall of size L) the 101010 solution is
rendered.  In a matrix form, the condition that determined $x^L_1$
($x_1$ for a given L) is:
\begin{equation}
\label{eq:8} \left( \begin{array}{c} 0 \\ x^L_1 \\ 1
\end{array} \right) = \left[ \begin{array}{ccc} -\frac{1}{\gamma}
& -1 & \frac{1}{\gamma} \\ 1&0&0 \\ 0&0&1
\end{array} \right]^L \left( \begin{array}{c}  x^L_1 \\ 0 \\ 1
\end{array} \right) = \mathcal{M}^L \left( \begin{array}{c}  x^L_1 \\ 0 \\ 1
\end{array} \right) .
\end{equation}
Where we assume symmetry of the soliton, so L must be even. In
other words, the condition that determine $x^L_1$ is:
\begin{equation}
\label{eq:9} \left( \left[ \begin{array}{ccc} -\frac{1}{\gamma} &
-1 & \frac{1}{\gamma} \\ 1&0&0 \\ 0&0&1
\end{array} \right]^L -\left[ \begin{array}{ccc} 0 & 1 & 0
\\ 1&0&0 \\ 0&0&1
\end{array} \right] \right)\left( \begin{array}{c}  x^L_1 \\ 0 \\ 1
\end{array} \right)=0
\end{equation}
Diagonalization of $\mathcal{M}$ is given by the matrix
$\mathcal{S}$:
\begin{equation}
\label{eq:10} \mathcal{S}^{-1} \mathcal{M} \mathcal{S} =
\mathcal{D}
\end{equation}
where
\begin{equation} \mathcal{D} = \left[ \begin{array}{ccc} 1 & 0 & 0
\\ 0& -e^{-i \theta} &0 \\ 0&0&-e^{-i \theta}
\end{array} \right] \end{equation}
and $\theta = \arctan(\sqrt{4 \gamma^2-1}) = \arccos(\frac{1}{2
\gamma})$.

\begin{figure}
  \includegraphics[width=7.7cm]{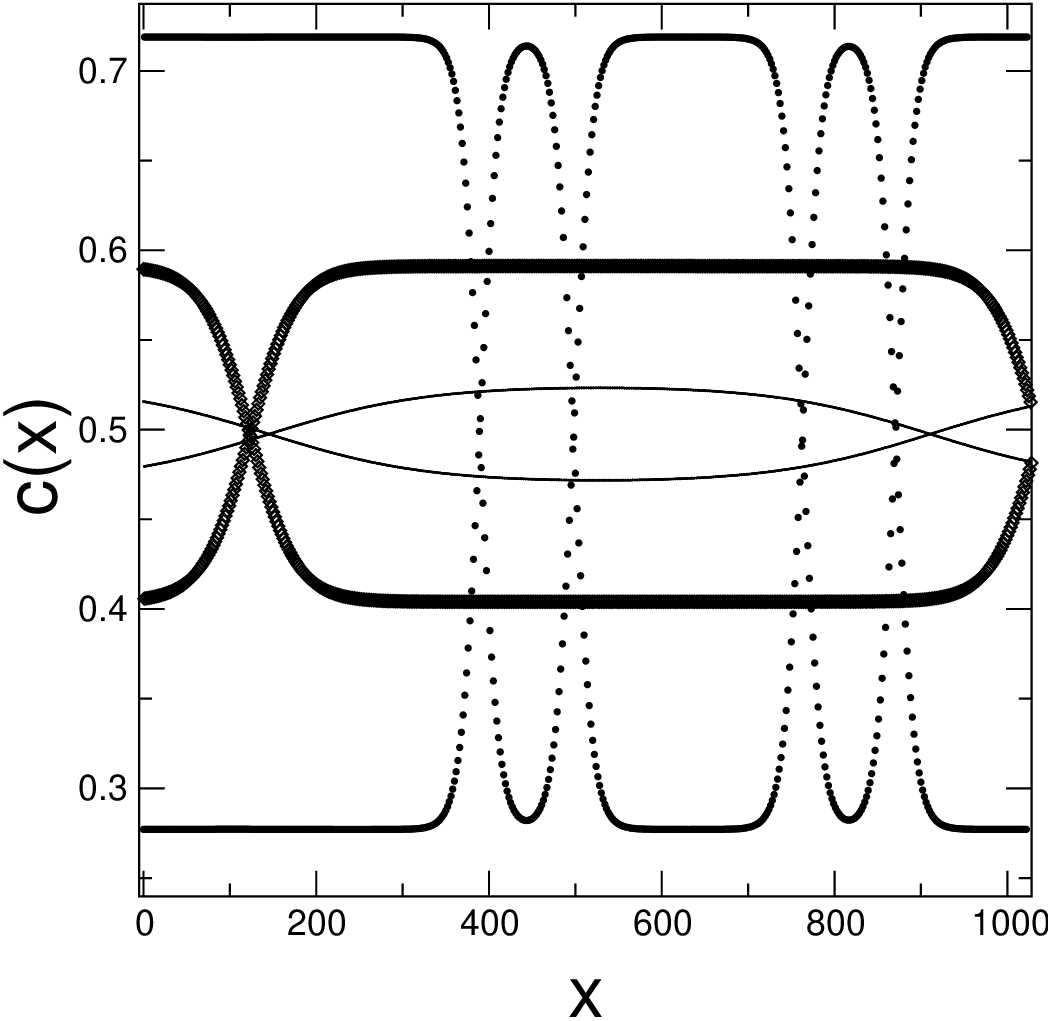}\\
  \caption{Solitons for nn interaction with $\gamma=0.505$ and $D=0.001$ (circles) $D=0.0012$ (heavy line)
   and $D=0.00124$ (line).
  $D_c = 0.00124378$  }\label{fig3}
\end{figure}

The eigenvalue problem (\ref{eq:9}) may be written in terms of the
diagonal matrix,
\begin{equation} x^L_1 = S_{11} - S_{12} \frac{r_{13}}{r_{23}} \frac{\cos(\frac{L
\theta}{2} + \varphi)}{\cos(\frac{L \theta}{2} + \eta)}
\end{equation}
and using $S_{12}=S_{13}^*=r_{12}e^{i \varphi}$ and
$S_{22}=S_{23}^*=r_{23}e^{i \eta}$ it implies that,
\begin{equation} x^L_1 = \frac{1}{2 \gamma +1} \left[ 1+
\sqrt{\frac{1-T_{L-1}(1/2 \gamma)}{1-T_{L+1}(1/2 \gamma)}} \right]
\end{equation}
where $T_n(x)$ is the n-th Chebyshev polynomial of the first kind.
After determining $x^L_1$ the same method may be used to derive a
general expression for all the elements of the size L soliton
($x^L_m$, where $1 \leq m \leq L$):
\begin{eqnarray} \label{soliton}
 x^L_m  =  \frac{1}{2 \gamma +1} \left[ 1 + (-1)^m
\sqrt{\frac{2 \gamma [1-T_{2m-1}(1/2 \gamma)]}{2 \gamma -1} }
\right.  \nonumber  \\ \left.
 -  (-1)^m \frac{2 \gamma x^L_1}{8 \gamma^2 -2}
\sqrt{1+T_{2m}(1/2\gamma)}   \right]
\end{eqnarray}
and it  fits perfectly the numerical experiment presented in Fig.
(\ref{fig2}, see captions).

The above analysis gives the shape of a soliton for any prescribed
length $L$,but simulation indicates that only \emph{one} soliton
size $L$ is selected for any set of parameters, and its length
diverges as $\gamma$ approaches its critical value. Looking
carefully at the solutions (\ref{soliton}) one realizes that all
other possible solitons admit  values for some of the $x_m$-s that
are either negative or larger than one, so these options are
unphysical (negative) or unstable to small perturbations.

The actual $L(\gamma)$ may be forecasted by a rough argument based
on a continuum approach. Defining the local deviation from the
one-zero solution, $c_n =1$ and $c_{n \pm 1} = 0 $,
\begin{equation}
c_{n \pm 1}= \delta_{n \pm 1} \qquad \qquad c_n = 1-\delta_n
\end{equation}
and plugging  it into Eq. (\ref{eq:7}) gives,
\begin{equation} \label{problem}
\delta_{n+1}+\delta_{n-1} = \frac{\delta_n}{\gamma}.
\end{equation}
Subtracting of  $2 \delta_n$ from both sides and taking the
continuum limit (i.e., assuming that the changes in $\delta$ from
site to site are small compared to $l_0$, here taken to be unity)
one gets,
\begin{equation} \label{simple}
\nabla^2 \delta(x) = - \frac{4 \epsilon}{1+2 \epsilon} \delta(x)
\end{equation}
with $\epsilon \equiv \gamma - \gamma_c$ goes to zero at the
transition, so $1+2\epsilon \approx 1$. The solution of  Eq.
(\ref{simple}) that satisfies the boundary conditions
$\delta(0)=0$ together with $\delta(L)=1$ is
\begin{equation}
\delta(x)=\frac{sin(2 \sqrt{\epsilon} x)}{sin(2 \sqrt{\epsilon})
L}.
\end{equation}
This expression  fails to converge smoothly to the "background"
ordered 010101 phase (at x=0 it has a finite slope), and is also
asymmetric. Put it another way, there are no nontrivial
heteroclinic orbits for a parabolic potential.  On the other hand,
close to the transition, where the size of the domain wall is
large, it seems that it should fit a solution of the continuum
approximation. The only way out
 is to pick a soliton size L such that the continuum
equation admit \emph{no solution at all}, i.e.,
\begin{equation}
L=\frac{\pi}{2 \sqrt{\epsilon}}. \end{equation} Such a choice
forces us back to the discrete equation (\ref{eq:7}) and its
solution (\ref{soliton}). This argument turns out to give the
right length of the stable soliton in the large $L$ ($\epsilon \to
0$) limit, as shown in Fig. (\ref{fig5}).

Let us consider now the domain walls for the finite diffusion
case.  As seen in Fig. (\ref{fig3}), there are also solitons for
the finite diffusion case, but now they admit tails that
asymptotically looks like $\delta \sim exp(-x/\xi)$, and $\xi$
diverges at the transition [e.g., at $D_c = \frac{2 \gamma -1}{4(2
\gamma +1)}$ for nearest neighbor interaction]. Defining a
vectorial  "order parameter" according to the larger $c$ values,
soliton solution interpolates between $(1,0)$ (larger $c$ on the
odd sites) to $(0,1)$ (even sites), and its shape is given by the
saddle point solution of the appropriate dynamics. Although the
determination of its full shape is difficult, it is possible to
determine $\xi$ by linearizing around one fixed point. For small
deviations, $c_n = A_0+A_{\pi/l_0}-\delta_n$ and $c_{n \pm 1} =
A_0-A_{\pi/l_0}+\delta_{n \pm 1}$ and Eq. (\ref{eq:7}) yields the
two coupled linear equations for odd and even n-s:

\begin{figure}
  \includegraphics[width=7.cm]{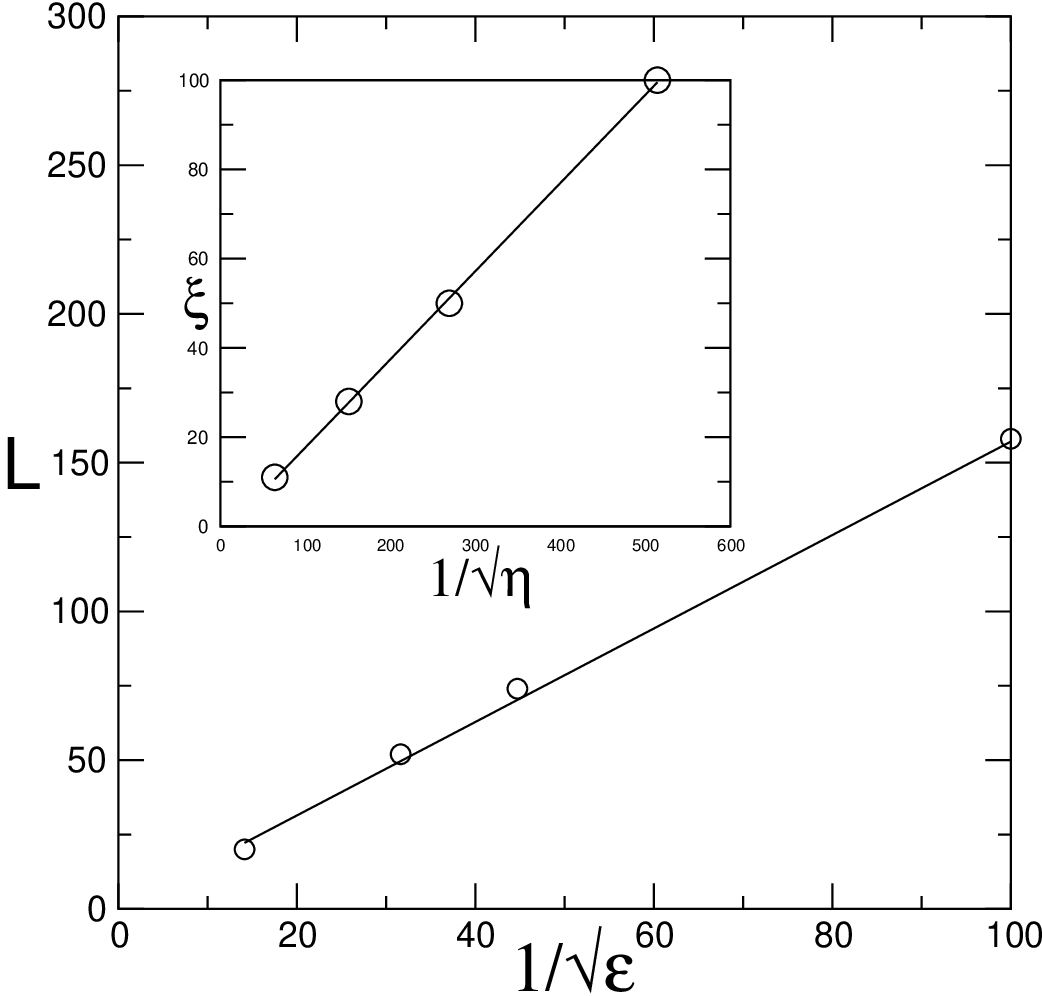}\\
  \caption{Soliton size ($L$)   as a function of $1/\sqrt{\epsilon}$,
   ($\epsilon = \gamma - \gamma_c$),
  for a nearest-neighbor
  interaction without diffusion. The circles are the results of a
  numerical simulation and the line is
  $\pi/(2 \sqrt{\epsilon})$. In the inset, the characteristic length of
  the soliton tail, $\xi$, is plotted against
  $1/\sqrt{\eta}$ for  $\gamma=0.505$ for the solitons shown in
  Fig. (\ref{fig3}). The circles come from the numerics and the
  line is the best linear fit that give a slope of 0.198, to be
  compared with $1/\sqrt{32} = 0.177$.} \label{fig5}
\end{figure}

\begin{eqnarray}
(2 s  + 2 d \gamma -1 + 2 D)\delta_n^{even} +(D- \gamma
s) (\delta_{n+1}^{odd} + \delta_{n-1}^{odd}) = 0    \\
(2 d  + 2 s \gamma -1+2D)\delta_n^{odd} +(D- \gamma d)
(\delta_{n+1}^{even} + \delta_{n-1}^{even})=0. \nonumber
\end{eqnarray}
where $s \equiv A_0 + A_{\pi/l_0}$ and $d \equiv A_0
-A_{\pi/l_0}$.  These coupled equation may be solved with the
ansatz $\delta^{even} \sim a_1 \exp(-x/\xi), \ \ \delta^{odd} \sim
a_2 \exp(-x/\xi)$ to give,
\begin{widetext}
\begin{equation} \label{xi}
\xi =  \left[ arccosh \left( \frac{1}{2} \sqrt{\frac{(4+64 D^2 -
32 D)\gamma^2+(20 D- 8D^2 -4)\gamma + 1-20 D^2}{D(\gamma - D- 2
\gamma D) }}.  \right) \right]^{-1},
\end{equation}
\end{widetext}
As the diffusion constant approaches its critical value, $D = D_c
- \eta$, $\xi$ diverges like $1/\sqrt{32 \eta}$. This prediction
is tested in the caption of Fig. (\ref{fig5}) against the numerics
and there is reasonable quantitative agreement, given the
difficulties in getting reliable numerical accuracy for the slope
of the logarithmic tail of the soliton.

\section{concluding remarks}

The model of logistic growth with long range interaction term may
serve as a generic, minimal model for competition for common
resource and pattern formation in excited media. In this paper
this model has been analytically discussed, with two main
outcomes. First, a general scheme for the identification of the
pattern forming instability has been presented, along with
explicit results for few common cases. Second, the defected
solutions for random initial conditions has been presented and
analyzed, and their characteristic length that diverges at the
transition is calculated.

The patterned solutions (like the 010101  for n.n. interactions)
are stable against small perturbation (like a low amplitude  white
noise). If instead of ...01010101.. one have ...01010(0.9)01.. the
0.9 site relaxes to 1. The domain walls, on the other hand, are
much less stable, and their density and dynamics has to be
strongly effected by an external noise. This is very much like the
situation in magnetic system, where the response functions of the
material are basically determined by the domain walls and not by
the "bulk". In magnetic systems, however, one may define the state
of the system at finite temperature ad a minimum of the free
energy function and consider the effect of noise simply as
temperature increase. The situation seems to be different for the
long range competition model: no simple Liapunov function exist
for this system, and the steady state is not derived from some
variational principle. In spite of that it is plausible to assume
that the defected solutions determine the response function of the
segregated phase, and maybe an effective dynamical equations for
the solitons may be set up to give an approximate Liapunov
function for this system.

\section{Acknowledgements}
 I thank Prof. David Kessler for  helpful discussions and
comments.

\end{document}